\documentclass[conference]{IEEEtran}
\IEEEoverridecommandlockouts

\usepackage{cite}
\usepackage{amsmath,amssymb,amsfonts}
\usepackage{algorithmic}
\usepackage{graphicx}
\usepackage{textcomp}
\usepackage{xcolor}
\def\BibTeX{{\rm B\kern-.05em{\sc i\kern-.025em b}\kern-.08em
    T\kern-.1667em\lower.7ex\hbox{E}\kern-.125emX}}
\begin{document}

\definecolor{jycklegreen}{RGB}{34, 139, 34}
\newcommand{\JP}[1]{\textcolor{jycklegreen}{[John: #1]}}

\title{Exploring Vibration-Defined Networking\thanks{John Pasquesi and Flavio Esposito are with the Department of Computer Science. Gianluca Davoli completed this work as a visiting scholar in the Department of Computer Science at Saint Louis University.  Jenna Gorlewicz is with the Parks College of Engineering, Aviation and Technology.}   \\[.75ex] \vspace{4mm}
  {\normalfont\large 
     John Pasquesi$^{*}$  \qquad Flavio Esposito$^{*}$ \qquad Gianluca Davoli$^{*\; \dagger}$ \qquad Jenna Gorlewicz$^{*}$%
  }\\[-0.5ex]
}

\author{
    \IEEEauthorblockA{%
        $^{*}$Saint Louis University, USA
    }
    \and
    \IEEEauthorblockA{%
        $^{\dagger}$University of Bologna, Italy
    }
}

\maketitle

\begin{abstract}
The network management community has explored and exploited light, copper, and several wireless spectra (including acoustics) as a media to transfer control or data traffic. Meanwhile, haptic technologies are being explored in end-user (wearable) devices, and Tactile Internet is being used merely as a metaphor. However, with rare exceptions and for smaller scoped projects, to our knowledge, vibration has been largely untouched as networking communication media.

In this paper, we share the lessons learned while creating and optimizing a pilot testbed that serves as an inexpensive starting point for the exploration of vibration-defined networking. We demonstrated the feasibility  (but not yet the scalability) of vibrations as a tool for a few network management mechanisms, such as resiliency, physical layer security, and as an innovative method for teaching networking concepts to individuals with visual impairments (VI). By demonstrating how vibrations could be programmable, we propose to the community a few open problems that could generate several applications. 
\end{abstract}


\section{Introduction}
An overarching goal of modern networks is softwarization, for policy programmability and adaptability. Aside from forwarding, many other network mechanisms have been redesigned with programmability in mind. From application mechanisms such as measurement~\cite{sdm},  network-level mechanisms such as scheduling~\cite{kkscheduling}, down to the physical layer with software-defined and cognitive radios~\cite{softran}, softwarization has touched nearly every aspect of modern networks. 

Additionally, an area that often warrants attention is the methods through which networking is taught. In recent years, a growing movement has brought focus to the need for making computer science education accessible to all individuals, particularly the blind and visually impaired (VI) communities. With many graphical and visual issues within networking, these individuals are often left behind.

With these two disparate issues in mind, we propose Vibration-Defined Networking (VDN) as an additional foundation for softwarization of the physical layer and as an innovative method for teaching networking topics. The remainder of this paper is an exploration of our VDN design and prototype implementation, and is organized as follows. The next section defines our contributions with respect to related work, while in Section~\ref{sec:arch} we dissect the general architecture of VDN.
Section~\ref{sec:testbed} covers the  experimentation and lessons learned from implementing the testbed, and explores the scalability of the testbed with a multi-hop system. In Section~\ref{sec:use} we cover use cases that we foresee as potential applications for VDN, and in Section~\ref{sec:future} we propose future research directions. 

\section{Related Work}

\subsection{Specialized vibration sensing}

There have been several studies that explored vibration sensing for specialized use cases. These articles have covered a range of interesting topics from termites using vibrations to select their food~\cite{Evans2005}, to binary telecommunication via cellular devices~\cite{Roy2015}. As an example of a specific application, Liu et al. ~\cite{Liu2017} used the vibrations from a simple finger touch on a surface to implement a virtual keyboard. While past literature has proposed many ways to send and measure vibrations,
we aim at providing an alternative form of networking communication, that can serve to transfer control traffic, or very low throughput data traffic.

\subsection{Out-of-channel networking}

A few other out-of-channel forms of communication have been proposed; from 60Ghz beams~\cite{cuttingthecord} to communication via light signals~\cite{Light-based-Positioning-Little} or power-line~\cite{powermanNSDI2018}, to the recent study involving Music-Defined Networking (MDN)~\cite{Hogan2018}. 
While acoustics may be an effective form of communication, it could be very unpleasant for humans to interfere with a MDN system, assuming that the human hearable spectrum is used.  
In our VDN design, we share some of the design principles from~\cite{Hogan2018} and believe that many of the proposed network management applications, as well as some of their limitations, also apply. Moreover, our physical layer propagation media is also flexible, opening new thought-provoking research and teaching directions and exciting (hidden) communication opportunities. The exploration of new communication media is also useful to expand potential applications, as seen in air-water communication~\cite{246296}.

\subsection{Teaching Using Vibration}

Touch is an important component of learning, in STEM disciplines for all students, but particularly those who heavily rely on touch as a primary communication channel ($e.g.$, blind and VI individuals). From hands-on learning experiences to the use of force feedback devices in virtual learning, there are many instances where touch has demonstrated its use in learning abstract concepts~\cite{Han2011,Lederman1987,Black2010, Tennison2019,L.Gorlewicz2019,Li2014}. Quorum, an evidence-oriented programming language, specifically has an ``auditory" track that enables individuals with VI to program ~\cite{Stefik2019,Ladner2017a,Quorum}. The testbed proposed in this work has the potential to contribute to this initiative, providing an avenue by which networking principles, which often employ visualizations, may be taught in a multisensory way --- catering to individuals with different learning styles or disabilities.
 
 \subsection{Advanced Communication Via Vibration}

This paper aims to use vibrations to communicate in a unique way. With this in mind, special attention should be given to the work of Roy et al (Ripple II)~\cite{Roy2016}. This work built upon previous efforts, proposed a few advancements in using vibration for communicating data ~\cite{Roy2015,Hwang2012}. The main two contributions of the Ripple II paper were the implementation of an OFDM-based vibratory radio using vibra-motors for transmission and microphones for receiving, and the creation of a completely functional system using vibration as a data communication method. Our paper aims to keep this work in mind, but our efforts are towards a self-contained, low-cost system, with goals specifically in the networking and education fields. Our work differs due to its focus on applications in network management and teaching rather than improving on the highly-involved vibration method, and in its use of piezoelectric discs as both a transmitter and receiver. Finally, we also differ as our design is oriented towards programmability of the vibration signal rather than merely another alternative way of communication. 
 
 \section{VDN Architecture Design}\label{sec:arch}

In this section we explore the mechanisms behind our VDN design, as well as of a management object model, that we implemented in support of vibration policy programmability. In the network management literature, an object is composed by the set of objects that we wish to manage, an API, to change the objects' attributes locally, and a management protocol.
By VDN policy, we mean a variant aspect of any of the mechanisms necessary to tune and adapt the vibration firmware to several application needs. 

\begin{figure}[t!]
\centering
\includegraphics[width=\linewidth]{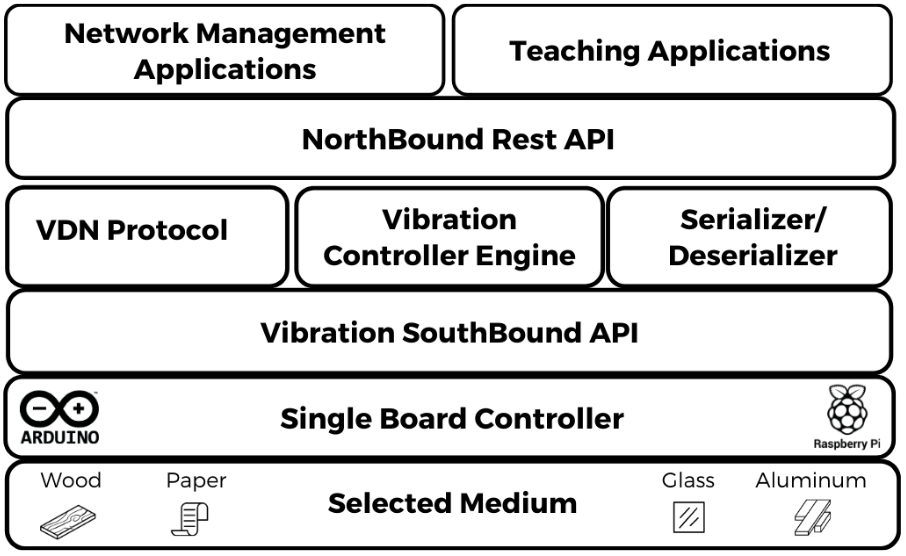}
\caption{Architecture overview: This architecture guides the format and design observed in the rest of the paper. This may be used as a reference for the overall system layout.}
\label{fig:arch}
\vspace{-4mm}
\end{figure}

The general overview of our architecture is illustrated in Figure~\ref{fig:arch}, and contains all key elements of the system. 
The physical apparatus, described in Section~\ref{sec:testbed}, is made up of the selected medium, a single board controller, and the vibration elements. 
The vibration interface includes the SouthBound API, a serializer/deserializer component, and the Vibration Controller Engine. 
Then, we have the VDN Protocol, for vibration policy programmability, the NorthBound Rest API, and the application logic. 
We give some examples of applications considering  a few use cases on Network Management and Teaching via vibration (Section~\ref{sec:use}).

\subsection{Rethinking Physical Layer Programmability}

Our communication model is centered around the idea of exploiting the properties of alternative physical layers; the selected medium may be any type of material that can effectively transfer vibrations. 
To control the hardware and transduce vibrations from the selected medium into network data (or control) layer-2 frames, we need a processing unit and a firmware unit; in our implementation, we have used a single board controller.
Such a hardware controller is responsible for interacting with the vibrations and the vibration SouthBound API, that logically sits on top of the firmware.

\subsection{Vibration interface}

The vibration interface allows the single board controller to communicate with the vibration controller engine and the NorthBound API. The SouthBound API is the single board controller's method of communicating with the vibration elements, and allows for sending and receiving vibrations. The sending of vibrations is possible via the Arduino's {\tt tone()} function, while receiving is enabled by the ArduinoFFT library, which performs a Fast Fourier Transform (FFT) on the incoming signal to determine the strongest frequency component. The Vibration Controller engine is a refined interface that allows for two key functions: {\tt vibration\_send()} and {\tt vibration\_receive()}. The {\tt vibration\_send()} function allows the user to send a specific frequency for a specific duration, while the {\tt vibration\_receive()} function waits for a signal and then returns the signal to the user. This vibration controller engine allows the VDN Protocol to easily send and receive signals.

\subsection{Networking protocol}

Arguing that out-of-channel signaling has the potential to provide support to network management operations, we designed a protocol to associate vibrations to network management events and tasks. The protocol can be used to assign signals to devices at the physical layer, and to agree on the medium access control policy or technique.  Moreover, the protocol is employed to map the assigned signals to the different network managements events, tasks, and applications. Finally, the protocol can potentially be used to encode network states (e.g., forwarding, routing, firewall) or for rapid (although not formal) network verification.

We implemented a few applications providing examples of usage of out-of-channel signals to support network management operations (Section~\ref{sec:use}): a path verification mechanism inspired by the well-known tool {\tt traceroute}, an application to help identify heavy-hitter traffic flows with vibrations, and one to help detecting potential Distributed Denial-of-Service attacks. In these toy examples, we partition VDN devices into two subsets: monitoring devices and collector devices. We connect each monitoring device to the data plane switch, configured to allow traffic mirroring on one of its ports. Every monitoring device runs the logic of our application and emits signals in response to specific events. By doing so, a collector device can sense monitoring signals and gather intelligence on specific network events. Such intelligence can in turn be used to modify network states, for example steering traffic by inserting new OpenFlow rules~\cite{openflow} when dealing with Software-Defined Infrastructures, or even simply invoking {\tt iptables} commands.

\begin{figure}[t!]
\centering
\includegraphics[width=\linewidth]{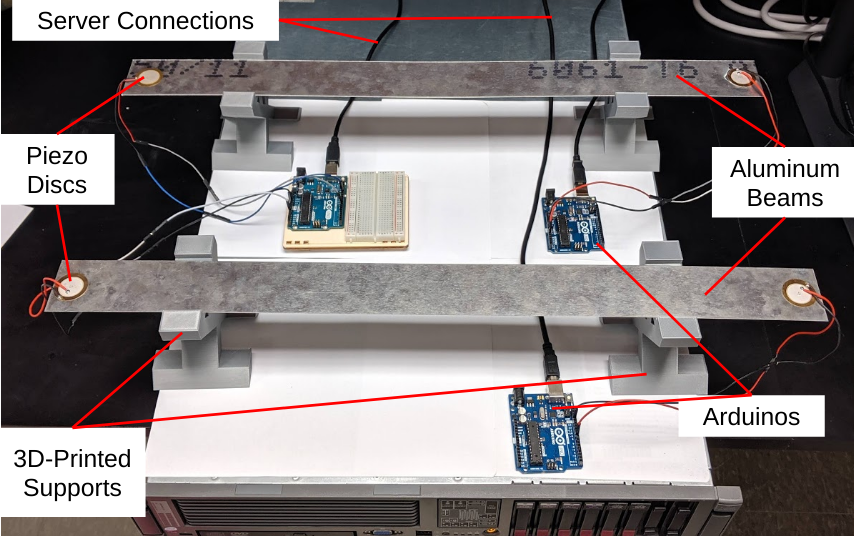}
\caption{Multi-hop Testbed: two of the base elements are combined to propagate the signal further. While in this image the two beams are parallel, note that this method could be used to extend the signal further along a single direction. This setup is similar to the single-beam apparatus, except that one arduino is connected to both input of one beam and output of the other.}
\label{fig:apparatus}
\vspace{-5mm}
\end{figure}

\section{VDN Testbed}\label{sec:testbed}

\subsection{Testbed Design}

For preliminary research into VDN, we constructed a pilot testbed to demonstrate the ability to communicate various signals effectively through vibration. Figure~\ref{fig:apparatus} shows the apparatus used for the multi-hop study. The testbed involved merely one of the beams with two Arduino boards. To facilitate the reproducibility of our results, we provide all our testbed details. An aluminum sheet of dimensions $620mm \times 50mm \times 1mm$  is supported by two 3D-printed structures that elevate the sheet $8cm$. We choose this sheet for its low cost and thin profile. We added the visible supports to enhance the propagation of vibration signals across the beams; their dimensions were not optimized. At both ends of the beam, a Luvay $27mm$ piezoelectric disc element (Model $\#$: Luvay000040) adhered to the top of the surface. Each piezo disc is connected to an Arduino Uno to control the sensing and send the vibration signals. We choose the combination piezo-Arduino for the sensitivity of the piezo as a sensor, and the ability of the Arduino to read analog signals.

\subsection{Investigations of the Testbed}
Three essential parameters affect system performance --- frequency of the drive signal, amplitude of the drive signal, and the properties of the physical medium. We ran three studies investigating each parameter in more detail in order to optimize the system.

\begin{figure}[t!]
\centering
\includegraphics[width=\linewidth]{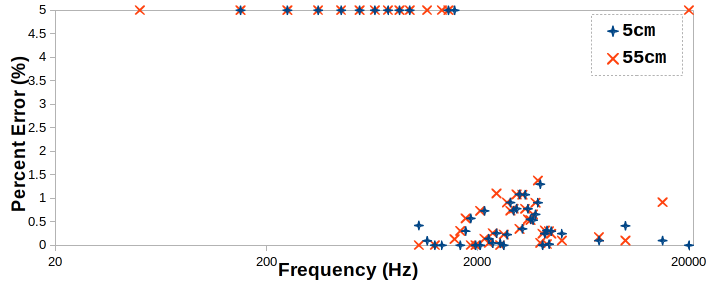}
\caption{Frequency of Vibration vs. Percent Error in Received Signal: The graph displays a percent error in the signal, capped at 5 percent for clarity. Note that the low error range for both signals coincide.}
\label{fig:freq}
\vspace{-2mm}
\end{figure}

Our first feasibility test sought to determine what range of frequencies we could send and receive across the beam. To determine such frequency range, we measure at $5cm$ and $55cm$ from the vibrating piezo disc, for a range of frequencies from $50Hz$ to $20,000Hz$. The samples were taken by connecting an oscilloscope to the receiving piezo and running an FFT on the input to determine the received frequency. The tested frequencies were in increments of $100Hz$ from $50-4500Hz$ and then at $5000, 7500, 10000,$, and $20000Hz$. We choose this scheme to give fine detail in the range the Arduino can read (up to $5000Hz$), and then the general trend above that frequency. Figure \ref{fig:freq} illustrates the results of this experiment. 

As can be seen in Figure \ref{fig:freq}, the lower frequencies ($<1750Hz$) are very difficult to detect. The most common causes of deviation from the expected signal were the detection of a harmonic rather than the exact frequency, or too small of amplitude to detect. 

The high-frequency signals are also challenging to detect, most likely because the piezo's resonance range is $4100-5100Hz$, and the farther away from this range frequencies are, the lower their amplitude. It is important to note that while the oscilloscope can read very high frequencies, the Arduino is limited to frequencies of $5000Hz$ or less, due to the speed of the analog to digital converter. From this test, we determined the optimal frequency range for our testbed was $1750-5000Hz$. We note that this range is characteristic of the selected hardware, and other ranges (such as lower frequencies) could be achieved with different vibration elements or processors. However, this experiment still demonstrates that even low-cost, readily available hardware has a functional range over which signals can be sent and received with $95\%$ accuracy.  

\begin{figure}[t!]
\centering
\includegraphics[width=\linewidth]{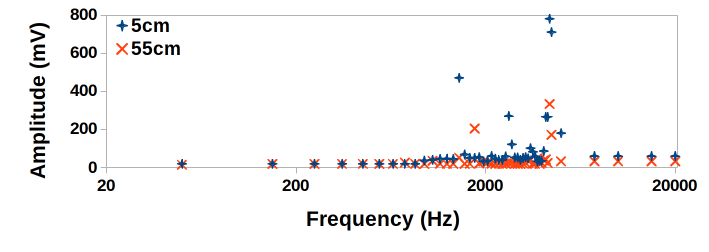}
\caption{Frequency of Vibration vs. Amplitude of Received Signal. Note that the amplitude at 5cm is consistently higher than at 55cm, as would be expected with the signal diminishing over distance. The range of higher amplitude frequencies also corresponds to the range determined in Figure \ref{fig:freq}. }
\label{fig:freq-amp}
\vspace{-2mm}
\end{figure}

Our second feasibility test sought to determine the range of frequencies that produced the most powerful signal from the piezo, $i.e.$, the signal with the most robust amplitude. To this aim, we ran the procedure detailed from the first experiment;  for this experiment, we measured the signal amplitude rather than the received frequency. The results of this experiment are presented in Figure~\ref{fig:freq-amp}.

Figure~\ref{fig:freq-amp} illustrates that at very low frequencies, the amplitude of the signal is too small to obtain meaningful data. At approximately $1750Hz$, the amplitude increases, which corresponds to the frequency range already determined in the previous experiment. This graph also illustrates that the strongest frequencies are around $1750, 3000,$, and $4500Hz$. At these frequencies, vibration signals are expected to travel further along a beam.

\begin{figure}[t!]
\centering
\includegraphics[width=\linewidth]{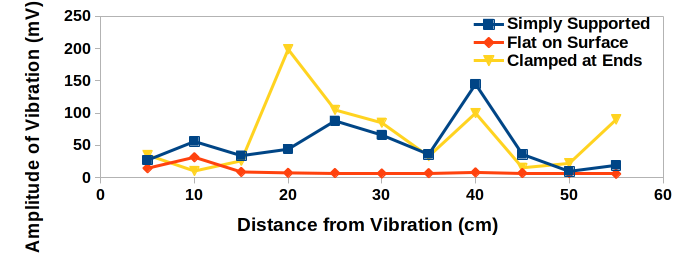}
\caption{Amplitude of Received Signal vs. Distance From Vibration: The comparison of amplitudes from 3 different boundary conditions. Note that the flat on surface method has the most rapidly decaying amplitude, while the simply supported and clamped at ends methods propagate the vibrations in patterns indicative of their ability to deform freely.}
\label{fig:bound}
\end{figure}

The final feasibility test sought to answer two questions: how does the length along the beam impact the signal, and how do the boundary conditions of the medium impact the signal? These questions are essential to understanding how the beams can be placed in practice --- both the distances they can span and how they can be attached to surfaces. To answer these questions, we sent a single frequency,  measuring the amplitude at points along the full length of the beam, ranging from $5cm$ away from the vibrating piezo to $55cm$ away in increments of $5cm$. We perform this experiment for three different boundary conditions: the beam supported at $5cm$ from each end (supported), the beam clamped to a support structure at each end (constrained at ends), and the beam laying flat on a lab bench (constrained throughout). Figure~\ref{fig:bound} displays the results of this experiment. 

This experiment demonstrates the importance of understanding how the signal propagates along the beam. When the beam is merely resting on the table, the signal dissipates quickly and remains at a low amplitude. Such low amplitude is due to the damping the table provides, which does not allow the signal to propagate easily. Meanwhile, the supported and clamped trials allow the beam to vibrate more freely, and thus the signal has peaks and dips as it travels along the beam. These conditions allow more reliable signals to be picked up further along the beam but also means that it is essential to understand where along the beam, the signal's amplitude fades. This understanding could provide a useful property: by strategically utilizing areas where the signal's amplitude is low, can we engineer a vibration system to prevent unwanted eavesdropping? 


\subsection{Lessons Learned from our Design}

Before arriving at the testbed configuration described above, we tested many combinations of equipment. This section briefly covers our attempts, and what we learned about communicating vibration effectively. The elements that were considered but not utilized include Raspberry Pis, MPU-6050 accelerometers, and various combinations of those elements with the Arduino and piezo. In preliminary testing, we encountered several issues with each of these attempts, and they lead to the following key lessons: sensor sampling rate is essential and needs to be tuned with care, the strength of the vibration is important, and more straightforward and inexpensive sensors performed reasonably well. 

The first key takeaway is the importance of the properties of the sensing element. A high sampling rate is essential. When we tested the accelerometer with the Raspberry Pi, its maximum sampling rate was $1000Hz$, which only allows frequencies up to $500Hz$ to be measured accurately. This frequency set does not cover the range of ideal frequencies the piezo can generate. The Arduino, on the other hand, has a sampling rate of $10,000Hz$, which allows for frequencies up to $5000Hz$ to be measured accurately. Thus, with what we know about the ideal range of frequencies on the piezo, a higher sampling rate, or a vibration element with stronger amplitudes at lower frequencies is necessary. 

A second key takeaway is the importance of the strength of the vibration signals. During our tests with the accelerometer, many vibrations received from the piezo were too weak to be detected by the accelerometer. However, when we used a more robust source of vibration (mobile phone), the vibration frequency was picked up accurately by the accelerometer. As amplitude decreased, the signal also became more likely to be misinterpreted. A more energetic vibration (or a high-resolution sensor) is necessary for reading signals. This information is something that may be used to improve future models, for example, by utilizing more potent vibration generators. However, due to their small form factors and ease of implementation, we believe using piezo elements as done in this work is desirable. 

Finally, simplicity is key. Both the Arduino and the Raspberry Pi are capable of sending vibrations through the piezo, but the Arduino code is a single line of code, while the Raspberry Pi requires more code and importing the GPIO library. Additionally, keeping all of the components connected to a single type of board reduces the complexity of the system. These design decisions not only make implementation and operation easier but also maintenance and future improvements to the system. 

Therefore, to reproduce or extend our results, we recommend beginning with a sensor with a high sampling rate, a strong vibration element, and as simple of an apparatus as possible.

\subsection{Multi-Hop Evaluation}

To evaluate the scalability of the apparatus to more considerable distances, we tested the multi-hop apparatus shown in Figure~\ref{fig:apparatus}. This exploration sought to determine the behavior of the signal (frequency and amplitude) when the signal was relayed to a second beam. This serves to demonstrate that if needed, the signal could be passed over greater distances. To determine the behavior of the signal, the frequencies $3250$, $3500$, $3750$, $4000$, $4250$, and $4500Hz$ were sent from the start of one beam, relay to another, and the amplitude and frequency of the final signal were measured. These frequencies were chosen for their strength as determined from the explorations of the testbed.

\begin{figure}[t!]
\centering
\includegraphics[width=\linewidth]{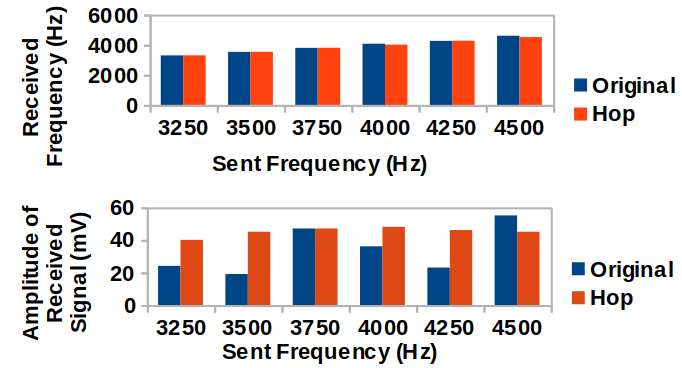}
\caption{Multi-hop Test Results. These results illustrate the values of frequency and amplitude at the receiving piezo of each beam- 'Original' for the first beam and 'Hop' for the second beam.}
\label{fig:multires}
\end{figure}

The results of this experiment, in Figure~\ref{fig:multires}, show that despite the varying amplitude, the frequency remains consistent after the hop. This result confirms that the signal can be extended.

\section{Use Cases}\label{sec:use}

\subsection{Reduce Attack Surfaces}

While it is still possible to have man-in-the-middle attacks in a VDN, we argue that the physical connection of the vibration elements and their properties makes it less likely to succeed.
To successfully collect information, send malicious signals or tamper with existing communication, an attacker would likely need to have a device physically connected to the vibration medium. At that point, the attacker would already have access to the server room, in which case network management traffic may not be the biggest problem. Even in other applications where physical access to devices is permitted to an attacker, touching the medium to feel and decode information is likely to alter it (unless a laser doppler vibrometer is used).

\subsection{Vibrations for Resiliency}

One of the key factors of any network is how well it handles failures or attacks. VDN has the potential to add to network resiliency, both actively and passively. 

Fate-sharing between the data and management plane is a serious issue~\cite{cuttingthecord}. Failures in the data plane network can cause damage to the management plane, preventing the execution of crucial tasks, including diagnostics and recovery. A VDN could help address this issue. With equipment regularly sending network management signals on a vibration plane, an interruption in the signals could notify of an issue without interrupting the normal flow of network operations. Essentially, a VDN serves as a new or additional out of channel method. 

Vibrations can also be used passively: by connecting a vibration sensor to a server fan or an hard drive so as to allow the system to detect irregularities in the vibration patterns, we can signal more serious failures before they occur. By preemptively warning of serious issues, the vibration plane could be a significant boost to the resiliency of the network as a whole.

\subsection{Vibration for Physical Layer Security}

Physical layer security (PLS) is a fairly recent solution that exploits the inherent differences between the properties of the channel between source and destination and the properties of the source-attacker or attacker-destination channels.
This technique has been shown to have major benefits but also drawbacks~\cite{Hamamreh2018}.
As signals propagating over wireless channels, also vibrations propagate differently when observed from different measurement points.
Vibrations also have the benefit of being dependent on the material on which they propagate; this, in turn, could allow security application programmers to customize the underlying propagation media, slightly or drastically. Moreover, surfaces can be engineered to enforce stronger attenuation in a given set of directions. Whether or not this property could help mitigate some of the security risks, for example a eavesdropper ability, is left as an open question. 

\subsection{Vibration as an Education Method}

Another fascinating application of VDN is its potential for teaching individuals with VI more about networking. There are many complex topics within networks that would prove difficult to teach to the VI community. The possibility of ``feeling" the network through VDN means that these individuals may be able to experience various concepts through vibration. For example, the idea of sending information within a network packet could be communicated by sending different patterns of vibration or varying the frequency significantly; vibrations can encode small pieces of information such as source or destination addresses, allowing students to feel, identify and learn by touching a packet. Similarly, the result of a checksum operation could be assigned to a specific vibration pattern, and examples given of when the checksum is intact versus when it is damaged could further clarify this notion. Computer Science Education for all is a large initiative, and many efforts have focused on making programming and programming languages more accessible. We believe this work has the potential to push this initiative forward, investigating how other facets of Computer Science education can be made more accessible and immersive for students with disabilities and diverse learning styles.

\section{Open Problems in VDN}\label{sec:future}

A potential application that we envision is the connection of several different nodes to one central point, where a single sensing node monitors several applications or the health of a full (perhaps small) networked system. This would of course require additional hardware, but might serve to provide a single unified point for network diagnostics. 

While our multi-hop vibration testbed demonstrates that is possible to send vibration signals over large distances, there are several challenged that need to be addressed. It would be interesting, for example, to explore how far the vibration could travel in a single hop. Exploring more precise sensors, or stronger vibration elements may also open up further possibilities. This idea also ties closely into the exploration of different media to propagate the vibrations. There may be other material that could propagate vibrations further.

\section{Conclusion}
In this paper, we have laid the foundation for Vibration-Defined Networking, and suggested potential uses of this (elsewhere explored) technology for physical layer security, to increase network resiliency and for inclusive educational purposes. To assess the practicality of our approach, we have built an architecture for vibration programmability and shared the experience obtained building several hardware testbeds. We analyzed the ability of different mechanical components to send and receive vibrations accurately.
We have exposed some limitations of our proof-of-concept prototype system, but also potential research directions.

\bibliographystyle{./IEEEtran}
\bibliography{./IEEEabrv,./vdn_noms}

\begin{thebibliography}{10}
\providecommand{\url}[1]{#1}
\csname url@samestyle\endcsname
\providecommand{\newblock}{\relax}
\providecommand{\bibinfo}[2]{#2}
\providecommand{\BIBentrySTDinterwordspacing}{\spaceskip=0pt\relax}
\providecommand{\BIBentryALTinterwordstretchfactor}{4}
\providecommand{\BIBentryALTinterwordspacing}{\spaceskip=\fontdimen2\font plus
\BIBentryALTinterwordstretchfactor\fontdimen3\font minus
  \fontdimen4\font\relax}
\providecommand{\BIBforeignlanguage}[2]{{%
\expandafter\ifx\csname l@#1\endcsname\relax
\typeout{** WARNING: IEEEtran.bst: No hyphenation pattern has been}%
\typeout{** loaded for the language `#1'. Using the pattern for}%
\typeout{** the default language instead.}%
\else
\language=\csname l@#1\endcsname
\fi
#2}}
\providecommand{\BIBdecl}{\relax}
\BIBdecl

\bibitem{sdm}
M.~Moshref, M.~Yu, R.~Govindan, and A.~Vahdat, ``Dream: Dynamic resource
  allocation for software-defined measurement,'' in \emph{Proc. of the 2014 ACM
  Conference on SIGCOMM}, 2014, pp. 419--430.

\bibitem{kkscheduling}
S.~G. Kulkarni, W.~Zhang, J.~Hwang, S.~Rajagopalan, K.~K. Ramakrishnan,
  T.~Wood, M.~Arumaithurai, and X.~Fu, ``Nfvnice: Dynamic backpressure and
  scheduling for nfv service chains,'' in \emph{Proc. of SIGCOMM '17}, 2017,
  pp. 71--84.

\bibitem{softran}
\BIBentryALTinterwordspacing
A.~Gudipati, D.~Perry, L.~E. Li, and S.~Katti, ``Softran: Software defined
  radio access network,'' in \emph{Proceedings of the Second ACM SIGCOMM
  Workshop on Hot Topics in Software Defined Networking}, ser. HotSDN
  '13.\hskip 1em plus 0.5em minus 0.4em\relax New York, NY, USA: ACM, 2013, pp.
  25--30. [Online]. Available: \url{http://doi.acm.org/10.1145/2491185.2491207}
\BIBentrySTDinterwordspacing

\bibitem{Evans2005}
\BIBentryALTinterwordspacing
T.~A. Evans, J.~C.~S. Lai, E.~Toledano, L.~Mcdowall, S.~Rakotonarivo, and
  M.~Lenz, ``{Termites assess wood size by using vibration signals},'' Tech.
  Rep., 2005. [Online]. Available:
  \url{www.pnas.orgcgidoi10.1073pnas.0408649102}
\BIBentrySTDinterwordspacing

\bibitem{Roy2015}
N.~Roy, R.~R. Choudhury, U.~Champaign, I.~Nsdi, N.~Roy, and R.~R. Choudhury,
  ``{Ripple: Communicating through Physical Vibration},'' in \emph{USENIX NSDI
  '15}, pp. 265--278.

\bibitem{Liu2017}
J.~Liu, Y.~Chen, M.~Gruteser, and Y.~Wang, ``{VibSense: Sensing Touches on
  Ubiquitous Surfaces through Vibration},'' in \emph{In Proc. of SECON 2017}.

\bibitem{cuttingthecord}
\BIBentryALTinterwordspacing
Y.~Zhu, X.~Zhou, Z.~Zhang, L.~Zhou, A.~Vahdat, B.~Y. Zhao, and H.~Zheng,
  ``Cutting the cord: A robust wireless facilities network for data centers,''
  in \emph{Proceedings of the 20th Annual International Conference on Mobile
  Computing and Networking}, ser. MobiCom '14, 2014, pp. 581--592. [Online].
  Available: \url{http://doi.acm.org/10.1145/2639108.2639140}
\BIBentrySTDinterwordspacing

\bibitem{Light-based-Positioning-Little}
\BIBentryALTinterwordspacing
E.~W. Lam and T.~D.~C. Little, ``Refining light-based positioning for indoor
  smart spaces,'' in \emph{Proceedings of the 4th ACM MobiHoc Workshop on
  Experiences with the Design and Implementation of Smart Objects}, ser.
  SMARTOBJECTS '18.\hskip 1em plus 0.5em minus 0.4em\relax New York, NY, USA:
  ACM, 2018, pp. 9:1--9:8. [Online]. Available:
  \url{http://doi.acm.org/10.1145/3213299.3213308}
\BIBentrySTDinterwordspacing

\bibitem{powermanNSDI2018}
\BIBentryALTinterwordspacing
L.~Chen, J.~Xia, B.~Yi, and K.~Chen, ``{PowerMan: An Out-of-Band Management
  Network for Datacenters Using Power Line Communication},'' in \emph{{NSDI}
  18}, Renton, WA, 2018, pp. 561--578. [Online]. Available:
  \url{https://www.usenix.org/conference/nsdi18/presentation/chen-li}
\BIBentrySTDinterwordspacing

\bibitem{Hogan2018}
\BIBentryALTinterwordspacing
M.~Hogan and F.~Esposito, ``Music-defined networking,'' in \emph{Proceedings of
  the 17th ACM Workshop on Hot Topics in Networks}, ser. HotNets '18.\hskip 1em
  plus 0.5em minus 0.4em\relax New York, NY, USA: ACM, 2018, pp. 155--161.
  [Online]. Available: \url{http://doi.acm.org/10.1145/3286062.3286085}
\BIBentrySTDinterwordspacing

\bibitem{246296}
\BIBentryALTinterwordspacing
C.~J. Carver, Z.~Tian, H.~Zhang, K.~M. Odame, A.~Q. Li, and X.~Zhou,
  ``Amphilight: Direct air-water communication with laser light,'' in
  \emph{17th {USENIX} Symposium on Networked Systems Design and Implementation
  ({NSDI} 20)}.\hskip 1em plus 0.5em minus 0.4em\relax Santa Clara, CA:
  {USENIX} Association, Feb. 2020, pp. 373--388. [Online]. Available:
  \url{https://www.usenix.org/conference/nsdi20/presentation/carver}
\BIBentrySTDinterwordspacing

\bibitem{Han2011}
\BIBentryALTinterwordspacing
I.~Han and J.~B. Black, ``{Incorporating haptic feedback in simulation for
  learning physics},'' \emph{Computers and Education}, vol.~57, no.~4, pp.
  2281--2290, 2011. [Online]. Available:
  \url{https://www.tc.columbia.edu/faculty/jbb21/faculty-profile/files/HantBlackCE2011-.pdf}
\BIBentrySTDinterwordspacing

\bibitem{Lederman1987}
\BIBentryALTinterwordspacing
S.~J. Lederman and R.~L. Klatzky, ``{Hand movements: A window into haptic
  object recognition},'' \emph{Cognitive Psychology}, vol.~19, no.~3, pp.
  342--368, jul 1987. [Online]. Available:
  \url{https://www.sciencedirect.com/science/article/pii/0010028587900089}
\BIBentrySTDinterwordspacing

\bibitem{Black2010}
\BIBentryALTinterwordspacing
J.~B. Black, ``{An embodied/grounded cognition perspective on educational
  technology},'' in \emph{New Science of Learning: Cognition, Computers and
  Collaboration in Education}.\hskip 1em plus 0.5em minus 0.4em\relax New York,
  NY: Springer New York, 2010, pp. 45--52. [Online]. Available:
  \url{http://link.springer.com/10.1007/978-1-4419-5716-0{\_}3}
\BIBentrySTDinterwordspacing

\bibitem{Tennison2019}
J.~L. Tennison and J.~L. Gorlewicz, ``{Non-visual Perception of Lines on a
  Multimodal Touchscreen Tablet},'' \emph{ACM Transactions on Applied
  Perception}, vol.~16, no.~1, pp. 1--19, 2019.

\bibitem{L.Gorlewicz2019}
J.~{L. Gorlewicz}, J.~{L. Tennison}, H.~{P. Palani}, and N.~{A. Giudice},
  ``{The Graphical Access Challenge for People with Visual Impairments:
  Positions and Pathways Forward},'' \emph{Interactive Multimedia [Working
  Title]}, pp. 1--17, 2019.

\bibitem{Li2014}
H.~Li and N.~A. Giudice, ``{The effects of 2D and 3D maps on learning virtual
  multi-level indoor environments},'' no. November, pp. 7--12, 2014.

\bibitem{Stefik2019}
A.~Stefik, R.~E. Ladner, W.~Allee, and S.~Mealin, ``{Computer Science
  Principles for Teachers of Blind and Visually Impaired Students},'' pp.
  766--772, 2019.

\bibitem{Ladner2017a}
R.~E. Ladner and A.~Stefik, ``{AccessCSforall},'' \emph{ACM SIGACCESS
  Accessibility and Computing}, no. 118, pp. 3--8, 2017.

\bibitem{Quorum}
\BIBentryALTinterwordspacing
``{The Quorum Programming Language},'' 2017. [Online]. Available:
  \url{https://quorumlanguage.com/}
\BIBentrySTDinterwordspacing

\bibitem{Roy2016}
N.~Roy and R.~R. Choudhury, ``{Ripple II: Faster Communication through Physical
  Vibration},'' in \emph{13th USENIX Symposium on Networked Systems Design and
  Implementation (NSDI 16)}, 2016, pp. 671--684.

\bibitem{Hwang2012}
\BIBentryALTinterwordspacing
I.~Hwang, J.~Cho, and S.~Oh, ``{Privacy-aware communication for smartphones
  using vibration},'' in \emph{Proceedings - 18th IEEE International Conference
  on Embedded and Real-Time Computing Systems and Applications, RTCSA 2012 -
  2nd Workshop on Cyber-Physical Systems, Networks, and Applications,
  CPSNA}.\hskip 1em plus 0.5em minus 0.4em\relax IEEE, aug 2012, pp. 447--452.
  [Online]. Available: \url{http://ieeexplore.ieee.org/document/6301465/}
\BIBentrySTDinterwordspacing

\bibitem{openflow}
\BIBentryALTinterwordspacing
N.~McKeown, T.~Anderson, H.~Balakrishnan, G.~Parulkar, L.~Peterson, J.~Rexford,
  S.~Shenker, and J.~Turner, ``Openflow: Enabling innovation in campus
  networks,'' \emph{SIGCOMM Comput. Commun. Rev.}, vol.~38, no.~2, pp. 69--74,
  Mar. 2008. [Online]. Available:
  \url{http://doi.acm.org/10.1145/1355734.1355746}
\BIBentrySTDinterwordspacing

\bibitem{Hamamreh2018}
J.~M. Hamamreh, H.~M. Furqan, and H.~Arslan, ``{Classifications and
  Applications of Physical Layer Security Techniques for Confidentiality: A
  Comprehensive Survey},'' \emph{IEEE Communications Surveys and Tutorials},
  vol.~21, no.~2, pp. 1773--1828, 2018.

\end{thebibliography}

\end{document}